\documentclass[preprint,10pt,a4paper]{aastex}
\usepackage[colorlinks,urlcolor=blue,citecolor=blue,linkcolor=blue]{hyperref}
\usepackage{amsmath}
\usepackage{graphicx}
\usepackage{multicol}
\newcommand{\maa}{\alpha \alpha} 
\newcommand{\mga}{\gamma \alpha}

\newcommand{\be}{\begin{equation}} 
\newcommand{\ee}{\end{equation}}
\newcommand{\nn}{\mbox{} \nonumber \\ \mbox{} }
\newcommand{\ba}{\begin{eqnarray}}
\newcommand{\ea}{\end{eqnarray}}
\newcommand{\om}{\omega}

\newcommand\eg{\textit{e.g.,\ }}

\newcommand{\Bf}{{magnetic field}}

\newcommand{\NS}{neutron star}
\newcommand{\NSs}{{neutron stars}}

\newcommand{\Lf}{Lorentz factor}

\begin{document}

\title{FBOTs and AT2018cow following electron-capture collapse of merged  white dwarfs}

\author{Maxim Lyutikov\\
Department of Physics and Astronomy, Purdue University, 
 525 Northwestern Avenue,
West Lafayette, IN
47907-2036  \\
and\\
Silvia Toonen\\
Anton Pannekoek Institute for Astronomy, University of Amsterdam, P.O. Box 94249, 1090 GE, Amsterdam}

\begin{abstract}
We suggest that fast-rising blue optical transients (FBOTs) and the brightest event of the class AT2018cow  result from 
 an electron-capture  collapse to a \NS\  following a merger of  a massive ONeMg white dwarf (WD) with  another WD.  Two distinct evolutionary channels  lead to the disruption   of the less massive  WD  during the merger and the formation of  a shell  burning non-degenerate star  incorporating  the ONeMg core. During the  shell burning   stage a large fraction of the envelope is lost to the wind,  while mass and angular momentum are added  to the core. As a result,    the electron-capture  collapse  occurs with a small envelope mass,  after  $\sim 10^2-10^4$ years. During the  formation of a  neutron star  as little as $\sim 10^{-2} M_\odot $ of the material is ejected at the bounce-off with mildly relativistic velocities and total energy $\sim$ few $ 10^{50}$ ergs. This ejecta becomes optically thin on a time scale of days - this is the FBOT.   During the collapse,  the   neutron star is spun up and magnetic field is amplified.
The ensuing fast magnetically-dominated relativistic wind from the newly formed neutron star shocks against the ejecta, and later against the wind.
The radiation-dominated forward shock produces the long-lasting optical afterglow, while the  termination  shock of the  relativistic  wind produces the  high energy emission   in a manner similar to Pulsar Wind Nebulae. 
%The late X-ray variability is akin to magnetically-driven Crab flares (and possibly late GRB flares).  
If the secondary WD was of the DA type,  the wind will likely have   $\sim 10^{-4} M_\odot$ of hydrogen; this explains the appearance of hydrogen late in the afterglow spectrum.
The  model explains many of the puzzling properties of FBOTs/AT2018cow: host galaxies, a  fast and light  anisotropic ejecta producing a bright  optical peak, afterglow high energy emission of similar luminosity to the optical,  and late infra-red features.
\end{abstract}


\section{Introduction} 

AT2018cow \citep{2018ATel11727....1S,2018ApJ...865L...3P,2018arXiv180800969P,2019ApJ...871...73H,2019ApJ...872...18M} is a mysterious astrophysical event.  It is likely to be the brightest member of the class of fast-rising blue optical transient  \citep[FBOT,][]{2014ApJ...794...23D}.  AT2018cow seems to be at a cross-road of supernova explosions (and associated complicated nuclear reactions, neutrino transports physics),  pulsars/magnetars,  (early) pulsar wind nebulae (PWNe), possibly GRBs and, as we suggest in the present paper,  the physics of white dwarf binaries.

AT2018cow had a few surprising features, the most important in our view being:
\begin{itemize}
\item the optical rise-time of $\leq 3$ days;  this is  the order of magnitude shorter than the conventional Ni-powered supernovae
\item the peak optical luminosity $\sim 4 \times 10^{44}$ erg s$^{-1}$; this exceeds the typical peak power of supernovae
\item the X-ray emission of initial power $\sim 10^{43}$ erg s$^{-1}$ had an extra component at $t \leq 15$ days, peaking at $\sim 40$ keV \cite[Fig. 6 in][]{2019ApJ...872...18M}
\item the  clear change of properties of the emission at $\sim$ 20 days \cite[Fig. 9 in][]{2019ApJ...872...18M}
\item an indication of the rising IR component at $t \geq 30$ days \cite[Fig. 5 in][]{2018arXiv180800969P}
\item the bright radio emission $t\geq 80$ day peaking at $\sim 10^{10}$ Hz \cite[Fig. 11 in][]{2019ApJ...872...18M}
\end{itemize}

These properties exclude normal Ni-powered supernovae and require a separate formation channel. We discuss one such possible channel in the present paper,  trying  to build  a coherent  model of AT2018cow. 

\section{The model:  a specific channel of WD mergers} 
In this paper we  discuss a scenario where fast-rising blue optical transients (FBOTs)   are powered by  the  electron-capture  collapse   following a merger of a massive ONeMg white dwarf (WD) with  another WD. Previously,  the  electron-capture  collapse was  mostly used in the  accretion models \citep{1976A&A....46..229C}, hence the name Accretion Induced Collapse, AIC  \citep[see also][]{1992ApJ...396..649T,1980PASJ...32..303M,1991ApJ...367L..19N,2006A&A...450..345K,2016A&A...593A..72J}. Some details of binary  evolution and of the collapse in such systems were previously discussed by \cite{2017arXiv170902221L}, see also \cite{2017ApJ...850..127B,2019MNRAS.484..698R,Yun17,2018A&A...619A..53T,2016MNRAS.458.3613S}.

Let us outline the main stages \cite[see][for a more detailed discussion]{2017arXiv170902221L}; also \cite{2019MNRAS.484..698R}. An initial   system with a primary mass   $M_1 \sim 6-10 \, M_\odot$  and a secondary mass   $M_2 \sim 3-6\, M_\odot$ forms  via two distinct evolutionary channels a double degenerate  CO-ONeMg  WD system. For  a sufficiently large mass ratio $q \equiv M_2/M_1 > q_{crit} \sim 0.25$  \citep{2004MNRAS.350..113M},  the  ensuing gravitational wave-driven     mass transfer is unstable, whereby  the less massive CO WD  is disrupted  on  a few orbital time scales  and  forms a  disk around the primary. \citep[Possible detonation of the CO WD secondary would  eject a small amount of mass, leaving the ONeMG core mostly intact][]{2018arXiv181100013K}. Disk accretion at high rates creates a spreading layer - a belt-like structure on the  surface of the primary \citep{1999AstL...25..269I,2009ApJ...702.1536B,2010AstL...36..848I,2013ApJ...770...67B,2016ApJ...817...62P}. After the spreading is complete on viscose time scale of $\sim 10^4$ seconds  \citep[\eg][]{2012ApJ...748...35S},  the resulting star of $\sim 2 M_\odot$ consists of a slowly rotating degenerate ONeMg core, and a  non-degenerate  envelope  rotating with a period of  hundreds of seconds. The non-degenerate  envelope expands  to $R_\ast \sim$ few $10^9$ cm. The star will emit near the Eddington limit and drive powerful winds.
 Angular momentum contained in the shell will be both lost to the wind and transported to the core through the (turbulent) boundary layer.

The merger product ignites shell CO burning, adding mass to the degenerate core; at the same time,  mass and angular momentum  are lost due to powerful winds.
For  an ONeMg  WD sufficiently close to the Chandrasekhar mass,  an electron-capture/accretion induced collapse   follows  after $\sim 10^2-10^4$ years.  
%The AIC proceeds inside-out, see Appendix  \S  \ref{collapse} - the core bounce is the first observed effect, see Appendix  \S \ref{bounce}.
%Core-bounce leads to an outgoing shock that  produces  a short  supernova-like explosion \citep[see also][]{1992ApJ...391..228W, 2006ApJ...644.1063D,2006A&A...450..345K,2010PhRvD..81d4012A,2012ApJ...747...88N,Fryer99}. 
 During the collapse,  the \Bf\ is amplified  \citep{2015Natur.528..376M},  and the \NS\ is spun to millisecond periods.
 
 To estimates the resulting \Bf\ and the spin frequency, we note that during  collapse of a core rotating initially with a spin frequency $\Omega$ and collapsing by a factor 
 $\eta_c \sim 100$ (from  a few thousand kilometers to a  few tens), the final \NS\ will rotate with $\om_c \sim \eta_c^2 \Omega$. The ratio $\om/\Omega =  \eta_c^2 \sim 10^4$ 
 also estimates by how much \Bf\ is twisted during the collapse.
 
 Thus, the final  toroidal  \Bf\  can be $\sim 10^4 $ times higher than the poloidal one. In addition,  the poloidal \Bf\ will be amplified by flux conservation. 
For example, if we start with $B_{WD} \sim 10^6$ G, flux conservation will give a factor $\eta_c^2 \approx 10^4$, while differential rotation will further boost that by 
$\sim 10^4$, reaching magnetar-like values of $B \geq B_Q$, $B_Q = m_e^2 c^3/(e \hbar)$.

  As a result,  the newly  born  spinning \NS\   will produce  a long lasting relativistic wind,  that  first shocks against the ejecta material and later on against  the wind material  lost during the  shell-burning stage. 
  The highly magnetized relativistic wind produced by a central \NS\ will interact with the fairly dense newly ejected material and dense pre-AIC wind,  producing an  $X$-ray afterglow at  the  {\it highly magnetized  reverse shock}, in a way similar to the case of afterglows from long GRBs, as suggested by  \cite{2017ApJ...835..206L}.

\begin{figure}[h!]
\centering
\includegraphics[width=.99\textwidth]{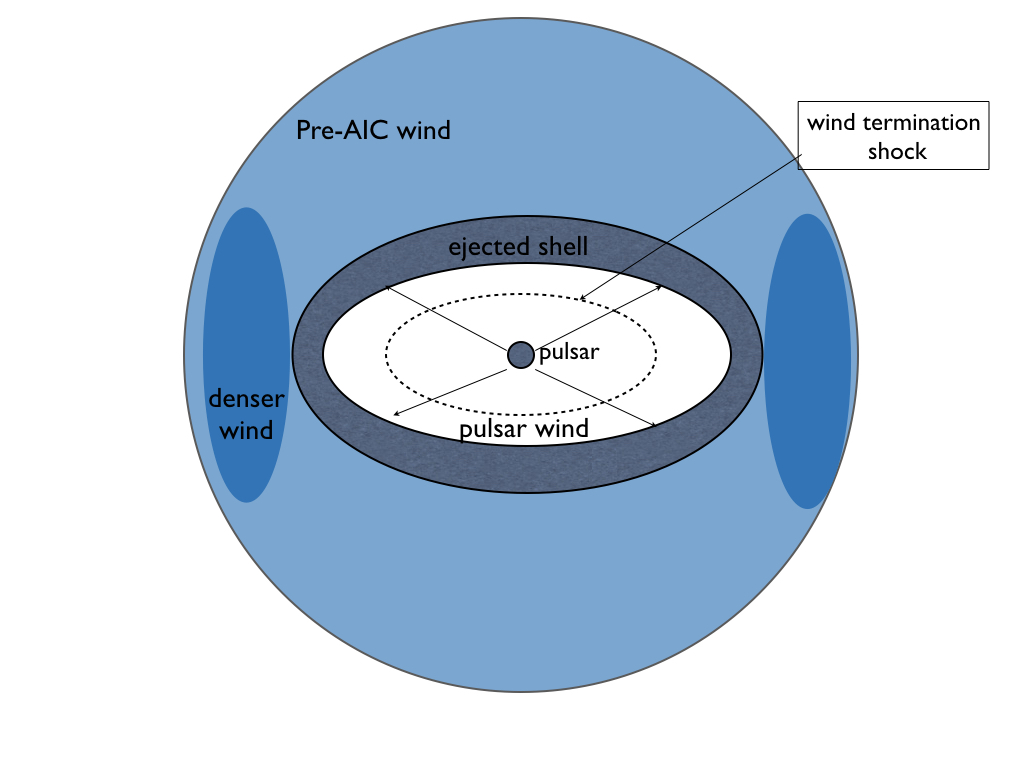}
\caption{Immediate surrounding of a FBOTs: the remaining NS generates  anisotropic pulsar-like wind (highly relativistic, highly magnetized with power $\sim \sin^2 \theta$), that interacts with the bounced-ejected shell, and the preceding  wind from the shell burning stage.  The pulsar winds,  the ejected shell and the pre-explosion wind are  all expected to be equatorially collimated.  }
\label{SurrondingSGRB} 
\end{figure}

 In Fig. \ref{SurrondingSGRB} we picture the immediate surrounding of an FBOT at times $\sim$ days-weeks  after the collapse. This picture is our  working model.

The key point of the model is that  a merger of a heavy ONeMg WD with another WD, and ensuring mass loss during the shell burning stage, results in a collapse of the core surrounded by a  fairly light envelope, tens of percent of $M_\odot$ at most. Depending on the particular collapse time one expects few cases when the AIC happens right before the shell burning is about to end.  This would  produce fast ejecta with small mass and leads to AT2018cow-like events. In other cases, envelopes of few tenths of solar masses are ejected, producing longer and less bright transients, which are still fast and luminous if compared with conventional supernovae.

 In addition to spinning up the core, a   large amount of  angular momentum of the shell (disrupted secondary WD) is  lost to the wind.
As a result the AIC can be direct, without formation of the accretion disk.  In this case the newly born \NS\  looses most if its rotational energy to the fast, relativistic wind.  For shorter shell-burning stage, there is too much angular momentum in the shell, that leads to the formation of the accretion disk, that spins down the central \NS\ in a propeller stage. As a result, a large fraction of the  energy is deposited into relatively slow, matter-dominated wind with low radiation efficiency.

%Regarding the origin of Type Ia SNe, our model is both somewhat independent of the active discussions on single versus double degenerate/mixed  origin of SNIa (in a sense that it can work in both cases), but  at the same time the model is closely related to the controversy. In any case, 
%As a working  assumption, we assume the double degenerate super-\Ch\ CO-CO WDs  scenario for SNIa.   the  merger of  ONeMg-CO WDs  will likely proceed in a  regime different to the more common CO WDs merger.

 \section{The pre-collapse wind and the ejecta}

\subsection{Pre-collapse wind}

 Let's assume that the post-merger/shell-burning star   launches a wind with velocity $V_w$. After time $t_w$ the edge of  wind reaches radius $r_{max} = R_\ast  + V_w t_w$. For a total mass in the wind $M_{env}$ we find 
\ba && 
\rho_w = \frac{\dot{M}}{4\pi r^2 V_w}
\nn &&
M_{env} = \int_{R_\ast } ^ {R_\ast  + V_w t_w}  \rho_w 4\pi r^2 dr = \dot{M} t_w
\nn&&
t_w = \frac{M_{env}}{\dot{M}}
\label{wind}
\ea
For mass loss rate of $\dot{M}=10^{-3} M_\odot$ yr$^{-1}$ and total mass in the envelope of $\sim 0.5 M_\odot$ this stage lasts $t_w \sim 5 \times 10^2$ yrs;  we adopt notation  $x_{n}= X/10^{n}$.

Optical depth to Thomson scattering trough the wind is $\sim 1 $ at 
\be 
r_{w}= \frac{\sigma_R \dot{M}}{4\pi m_p V_w}= 2\times 10^{13}   \dot{m}_{-3}   V_{w,8}^{-1}{\rm cm}
\label{rww}
\ee

In fact, the wind is composed of the WD material rich in heavy elements. Thus, it's opacity is similar to the case of type Ia supernova - it is dominated by line transitions, in expanding wind and can be an order of magnitude higher than free-free scattering \citep{1977ApJ...214..161K,2000ApJ...530..757P,1998ApJ...495..617H}. 

\subsection{Ejecta }

As the  shell is  accreted onto the proto-neutron star, a narrow outer  layer will be ejected.  Studies of the AIC ejected mass
 predict $\sim 10^{-3} - 10^{-1} M_\odot$ ejected \citep{1992ApJ...391..228W,Fryer99,2010PhRvD..81d4012A,2019arXiv190408427S}.
%,   see Appendix  \S  \ref{bounce} for a simple consideration of the collapse of a polytropic star.

 These studies were mostly concerned with single degenerate scenario of AIC, basically with no envelope. It is not clear how a presence of a  tenuous envelope, of few $0.1M_\odot$, would affect the ejecta mass. 
Let us assume, for a particular case of AT2018cow,  a bounced  ejected mass of  $M_{ej}$, with maximal velocity $V_{ej,0}$. 
For homologous expansion with $ v \propto r$, the energy in the ejected part is
\be
E_{ej} = \frac{3}{10} M_{ej}V_{ej,0}^2
\ee
 Before the ejecta slows down due to the interaction with the pre-existing wind, its density evolves according to 
\be
\rho_{ej} = \frac{3}{4 \pi} \frac{M_{ej}}{(V_{ej,0} t_{ej} )^3} 
\label{rhoej}
\ee
where $V_{ej,0}$ is the maximum velocity of the ejecta. 

For numerical estimates we chose  $M_{ej} \approx  10^{-2} m_{ej, -2} M_\odot$.  (Many of the final relations  depend weakly on  ejecta mass, $\propto M_{ej} ^{1/4}$, \eg\ Eq. (\ref{R}) and \ref{B1})). The velocity of the ejecta will be related to the escape velocity from the proto-neutron star,
\be
V_{ej,0} \approx \sqrt{ \frac{G M_{NS}}{R_{proto-NS}}} = 7 \times 10^9 {\rm cm s}^{-1}= 0.26 c
\ee
for a proto-\NS\ radius of  $R_{proto-NS}=30$km.
The ejecta energy is then
$
E_{ej} \approx 4\times 10^{50} m_{ej,-2}$ erg.
Thus, our model naturally produces fast and light ejecta.

\subsection{Ejecta-wind interaction } 

The ejecta will launch a forward shock into the pre-explosion wind. As the shock propagates through the wind the accumulated mass is then
 \be
 M_{acc} \approx \dot{M} t \frac{V_{ej,0}}{V_w}
 \ee
 It becomes equal the ejecta mass at
 \be 
 t_{slow} = \frac{M_{ej}}{\dot{M}} \frac{V_w}{V_{ej,0}} \rightarrow 4 \times 10^6 m_{ej, -2}   V_{w,8} \dot{m}_{-3}^{-1}sec
 \label{tslow}
 \ee
Thus, the external wind has little effect on the ejecta until  approximately a month after the explosion. Before that the ejecta is in free expansion stage.

This will be approximately the time that the NS-driven shock exits the ejecta and enters the wind. We associate this with the transition of the afterglow properties at $\sim 20$ days \citep{2019ApJ...872...18M}.

\section{Optical transient}

 \subsection{Optical emission-I:  expansion of the ejecta}

As discussed by \cite{2014ApJ...794...23D}, fast optical transients can be powered either by the thermal energy of a  low mass ejected envelope or by the central engine.
In the first case
The Thomson optical depth  through  ejecta is 
\be
\tau_{ej} = \frac{\rho_{ej} }{m_p} \sigma_T V_{ej,0} t = 
 \frac{3}{4 \pi} \frac{M_{ej}  \sigma_T }{m_p (V_{ej,0} t)^2}  =  3 m_{ej, -2}  t_{d}^{-2}
 \label{01}
 \ee
where $t_d$ is time in days since the explosion.
The surface $\tau =1$ evolves with time according to 
\be
r_{ej, \tau=1}= V_0 t \left( 1 - \frac{2\pi}{5}  \frac{m_p V_0^4}{ E_{ej} \sigma_T} \right)
\ee
It reaches maximum at
\ba &&
t_{ej, max}= \sqrt{ \frac{5}{6\pi}} \sqrt{ {E_{ej} \sigma_T}{m_p V_0^4} }\approx 1 m_{ej, -2}^{1/2} {\rm day} 
\nn && 
r_{ej, max}= \sqrt{ \frac{10}{27\pi}}\sqrt{ {E_{ej} \sigma_T}{m_p V_0^2} } \approx 5 \times 10^{14} m_{ej, -2}^{1/2} {\rm cm}  
\label{rej}
\ea
 This explains the short rise time of the transient. Also note that radius (\ref{rej}) is larger than the radius  (\ref{rww}) when the wind becomes optically thin. Thus, the maximum of the ejecta emission is not affected by scattering in the pre-explosion wind.

In fact, for the expanding optically thick plasma the peak  in luminosity  will appears before the shell becomes optically thin due to the fact that photons diffuse out somewhat faster  \citep{1982ApJ...253..785A}. The peak time is 
\be
t_{pk} \approx \sqrt{ \frac{M_{ej} \kappa}{4\pi V_{ej,0} c} }\approx 0.4 m_{ej,-2} {\rm days}
\label{02}
\ee
where $\kappa \approx 0.1 $ cm$^2$ g$^{-1}$ is an estimate of the effective
opacity due to electron scattering. Estimates (\ref{01})  and  (\ref{02}) explain the short rise time of the optical light curve.

The free-free emission of the optically thin part of the  ejecta is fairly small,
\ba &&
L_{ej, ff} \approx j_{ff} n_{ej}^2  (V_{ej,0} t)^3 =4 \times 10^{39} {\rm erg\, s}^{-1}  m_{ej, -2}^2 T_4^{1/2} t_d^{-3}
\nn &&
j_{ff}=2.4 \times 10^{-27} \sqrt{T} n^2 
\label{jff}
\ea
where $j_{ff}$ is free-free emissivity \citep[][]{1999acfp.book.....L}. 
Thus, we associated the early  fast optical transient with the emission of the ejected shell.

 \subsection{Optical emission-II:  forward shock from the NS-driven wind}
 \label{optical}
 
 The NS wind is  shocked at the termination shock and will also produce a forward shock. Similar amount of energy will be dissipated in both shocks. Particles accelerated at the termination shock  produces  the non-thermal X-ray emission, while the forward shock will produce the long lasting  optical emission (in addition to the emission produced by the ejecta).
 
 Let us first consider the forward shock emission.
 The forward shock initially propagates through the ejecta, and later-on through the pre-explosion wind. 
%  \subsubsection{Propagates through the ejecta - radiation-dominated shock at early times}
The shock driven by the NS wind will be modified by radiation pressure,  \citep{1976ApJS...32..233W,2010ApJ...725...63B,2010ApJ...724.1396O,2018MNRAS.477..816L}. The observed properties of mildly relativistic shocks  are fairly complicated and often have steep dependence on the underlying parameters  due to steep dependence of photon production rates, phone escape and pair production on the plasma properties. 
Qualitatively, radiation-dominated shocks can temporarily  reach temperature exceeding the shock jump conditions (even taking into account radiation pressure). For fast photon production this may results in isothermal jumps \citep{LLVI,2018MNRAS.477..816L}, in which case the peak temperature exceeds the final temperature by a factor of a few. For slow photon production the temperature peak may exceed the final temperature by a large amount \citep{2018MNRAS.474.2828I,2018arXiv181011022I}. 

Let us  give  here an order-of-magnitude estimate of the {\it final} post-shock temperature in strongly radiationally-dominated shock.
  The post shock pressure is 
 \be
 p_{FS} \sim \frac{L_w}{4\pi R_{PWN}^2c} = \frac{\sqrt{3}}{16 \pi} \frac{\sqrt{ L_{w,0}  M_{ej}}}{\sqrt{c} V_{ej,0}^{3/2} t^{5/2}}
 \ee
 The post-shock pressure is contributed  both by matter pressure, $\sim n T$ and radiation pressure $\sim 4 \sigma_{SB} T^4/c$. Radiation pressure dominates for  \citep{1976ApJS...32..233W,2010ApJ...725...63B,2010ApJ...724.1396O,2018MNRAS.477..816L} 
 \be
 V_s \gg \frac{ (\rho \lambda_C^3)^{1/6} c}{ m_e \mu^{2/3} }\approx 10^{5}  m_{ej,-2}^{1/6} t_d^{-1/2} {\rm cm\, s}^{-1}
 \label{vs1}
 \ee
 where $\lambda_C =\hbar /(m_e c)$ and $\mu = m_p/m_e$. Comparing with (\ref{VS}) we conclude that the forward shock is 
 radiation-dominated   with post-shock temperature (far downstream!)
 \be
 T_{FS} \approx \left( \frac{c L_{w,0}  M_{ej}}{\sigma_{SB}^2V_{ej,0}^3 t^5 } \right)^{1/8}=
 4\times 10^4 \,m_{ej,-2} t_d^{-5/8} {\rm K}
 \label{TFS}
 \ee
 This matches the observed temperatures at early times, both in value and (presumably)  the temporal decrease \citep{2019ApJ...872...18M}.
 
 %We may also verify that the free-free photon production rate $\dot{n}_{ph} \sim j_{ff}/( n T)$,  (emissivity  (\ref{jff}) divided by the number density $n$ and  by photon energy $\sim T$) is higher than the dynamical time,  \be \dot{n}_{ph} t \approx 10^2 m_{ej,-2} T_{K,4}^{-1/2} t_d^{-2} \ee  for approximately 10 days after the explosion.

% \subsection{Later forward shock emission}

 On the other hand, radiation-dominated shocks require sufficiently high optical depth, at least of the order of $\tau \sim c/V_s$, while final stationary configurations may be reached at optical depths of thousands \citep{2018arXiv181011022I}.  This is not really satisfied in the particular case: 
from $r \sigma_T n \sim c/V_s$,  with $V_s$ given by (\ref{VS}) and density by (\ref{rhoej}), the condition  $\tau \sim c/V_s$ is satisfied 
for
 \be
 t \leq \frac{ L_{w,0}^{1/3}  M_{ej} ^{1/3} \sigma_T^{2/3}}{ m_p^{2/3}  c V_0} = 0.1  m_{ej,-2}^{1/3}\, {\rm days}
 \label{tb}
 \ee
 Thus, only at very early times the shock is highly radiation-dominated.
 
At the moment of shock breakout, at $t \leq 1 $ day, we expect an X-ray flash with duration $\sim 10^4$ seconds, Eq. (\ref{tb}) and luminosities $\sim 10^{42}-10^{43}$ erg s$^{-1}$ \citep[\eg][]{2014ApJ...788..113S}.
  Later on, after the shock is no longer radiation dominated, the post-shock temperature evaluates to 
 \be
 T_s = \frac{3}{16} m_p  V_s^2 \approx \frac{m_p L_{w,0}^{1/2} V_0^{3/2} \sqrt{t}}{ \sqrt{c M_{ej} }}=5 \times  10^5  t_d^{1/2}  m_{ej,-2}^{-1/2} {\rm eV}.
 \ee
%  This is much higher than the observed temperatures of $\sim$ few eV.
 The corresponding free-free luminosity is far too small $\sim 10^{35}   m_{ej,-2}^{7/4} t_d^{-11/4}$  erg s$^{-1}$.

\subsection{Anisotropy}

Thus, optical emission is  puzzling - it is hard to see how requirement of radiation-modified shocks (and hence large optical depth) can be reconciled with short transient time scales (and hence small optical depth). A possible answer is anisotropy. All the ingredients - pre-collapse wind, ejecta and the NS wind are expected to be anisotropic, see Fig. \ref{SurrondingSGRB}.
First, the wind is launched by star that rotates with nearly critical velocity. Even Solar wind is highly anisotropic: slow and dense in the equatorial sector and fast in the polar regions (with approximately constant $d \dot{M}/d\Omega$). Second, the newly formed NS can, under certain parameters, be nearly critically spinning, so that the ejecta is anisotropic as well \citep[see][]{2017arXiv170902221L}. Third, the remaining NS's wind is  equatorially collimated,  with power $\propto \sin^2 \theta$ \citep{1973ApJ...180L.133M}.

There are strong observational arguments in favor of anisotropy. First, hydrogen and helium lines show spectral asymmetry,  with a  tail towards longer wavelengths \citep{2019ApJ...872...18M}. This can be explained if the line of sight samples unevenly the ejecta.

Anisotropic ejecta can also reconcile the requirements of low ejecta mass, $\sim 10^{-2} M_\odot$ (and hence early transparency) and the requirement of larger ejecta mass, to keep the NS wind-generated forward shock to be radiatively dominated, which would nicely explain the optical temperature, Eq. (\ref{TFS}).

An alternative possibility is that the later optical emission originates as a synchrotron emission from particles accelerated at the forward shock. One can construct a model of particle acceleration at the forward shock following the standard GRB parametrization \citep[\eg][]{piran_04}, with $\epsilon_B \sim 10^{-4}$, $\gamma \sim 10^3$, in the fast cooling regime (fast cooling for the forward shock is important - X-ray luminosity for the termination shock  and  optical from the forward shock are comparable,  \citep{2019ApJ...872...18M}. But the expected spectrum will be non-thermal.

 \section{Non-thermal emission in FBOTs:  pulsar-like  termination shock in fast  NS wind}

\subsection{Wind power}

The newly created \NS\   (NS) is spun up to short periods and \Bf\ is amplified.
The central NS will  produce a 
 highly magnetized wind  that shocks against the ejecta (and,  later on,  against the pre-explosion wind). The NS wind-ejecta interaction will produce two shocks: forwards shock in the ejecta and termination shock in the wind.
 It is the wind termination shock that produces the X-ray emission, while the radiation-dominated forward shock produces the optical transient, see \S \ref{optical}. 
  In the termination shock the accelerated particles will produce synchrotron emission  in the fast cooling regime, so that a large fraction of the wind power will be emitted as radiation. \citep[See][for discussion of emission produced at the highly magnetized termination shock in GRBs]{2017ApJ...835..206L}. 
 
 In the fast cooling regime most of the power given to the accelerated particles is emitted. 
Let us then identify the observed  X -ray luminosity with NS  wind power. It is nearly  constant at $L_0 \approx 10^{43}$ erg s$^{-1}$ 
  until $t_\Omega \sim$ 20 days and then falls  $\propto t^{-2}$ \citep[Fig. 5 in][]{2019ApJ...872...18M}. Using the  initial pulsar spin-down power and spin-down time $t_\Omega$, we find the \Bf\ and the initial spin
  \ba &&
  B_{NS}\approx  \frac{ c^{3/2} I_{NS}}{\sqrt{L_0} R_{NS}^3 t_\Omega} =8 \times 10^{14} \, {\rm G}
 \nn &&
 \Omega_0 =\sqrt {\frac{L_0 t_\Omega}{I_{NS}}} =140 \, {\rm rad s}^{-1}
 \label{21}
\ea
At $t> t_\Omega$ we have $L \propto t^{-2}$.
Thus we suggest that a magnetar-type object is formed; its initial spin is not very high - $45$ milliseconds.
The wind power at time $t$ is then
\be
L_w = \frac{L_{w,0}} {(1+t/t_\Omega)^2}
\label{Lw}
\ee
(In fact, since observed X-ray luminosity  is a lower limit on wind power, estimates (\ref{21}) are upper limit on \Bf\ and lower limit on the initial spin.

The dynamics of shock driven by the wind with power (\ref{Lw}) will depend on two factors: spin-down time $t_\Omega$ and whether the shock is in the ejecta, \S  \ref{11}, or in the pre-explosion wind, \S \ref{112}. We consider the two cases next.

\subsection{Propagation of NS wind-driven shock through  ejecta }
\label{11}

The newly formed \NS\ generates powerful wind that first propagates within the ejecta, and later on through the  pre-explosion wind.
Let us consider the dynamic of the wind-driven shock propagating through ejecta with density (\ref{rhoej}). In the  Kompaneets approximation \citep{1960SPhD....5...46K,1995RvMP...67..661B} \footnote{Kompaneets approximation assumes supersonic driving, while Sedov scaling assumes subsonic driving of the expansion. At early times, when the termination shock is close to the contact discontinuity, the  Kompaneets approximation  is more justified that the Sedov's.},  the relativistic wind with power $L_w$ will produce a cavity expanding according to
 \be
 \frac{L_w}{4\pi R^2c} =\rho_{ej} \left( \partial_t R - \frac{R}{t} \right)^2
 \label{Komp}
 \ee
 Given the wind power (\ref{Lw}) and  density (\ref{rhoej}) the radius of the cavity evolves according to
 \ba && 
R_{PWN}\approx  \frac{ (L_{w,0} t_\Omega)^{1/4} V_{ej,0}^{3/4} }{( c M_{ej})^{1/4}}
t  \sqrt{\arctan \sqrt{t/t_\Omega}}=
\left\{ \begin{array}{cc}
 \frac{ L_{w,0}^{1/4} V_{ej,0}^{3/4} }{( c M_{ej})^{1/4}}  t^{5/4},& t \rightarrow 0\\
\sqrt{\frac{\pi}{2}}  \frac{ (L_{w,0} t_\Omega)^{1/4} V_{ej,0}^{3/4} }{( c M_{ej})^{1/4}}   t ,& t \rightarrow \infty
\end{array}
\right.
\label{R}
\ea
(function $ x \sqrt{\arctan \sqrt{x}}$  has limits $x^{5/4}$ for $x \ll1 $ and $\sqrt{\pi/2} x$ for $x \gg 1$). 
For example, at one day the \NS\ produces a cavity if a size $R \approx 6 \times 10^{13} m_{ej,-2}^{-1/4}$.

The relative velocity of the shock with respect to the ejecta is
\be
V_s = 0.37 \frac{ ( L_{w,0} t)^{1/4}V_{ej,0}^{3/4}  }{( c  M_{ej})^{1/4}} = 2 \times 10^8 \, t_d ^{1/4} m_{ej,-2}^{1/4}\, {\rm cm \, s}^{-1}
\label{VS}
\ee
Thus, it changes only slowly with time.

The corresponding equipartition post-termination shock \Bf\ (in the highly magnetized wind) is
\be
B= \frac{ \sqrt{2 L_w}}{\sqrt{c} R} = 
\sqrt{2} \frac{ L_{w,0} ^{1/4} M_{ej}^{1/4} }{ c^{1/4}  V_{ej,0}^{3/4}}  \left( t^{5/4} \sqrt{1+t/t_\Omega} \right)^{-1} \approx
300 {\rm G} m_{ej,-2}^{1/4}  \left( t_d^{5/4} \sqrt{1+ 0.05t_d} \right)^{-1}
\label{B1}
\ee
($t_d$ is time measured in days.) Thus, at early times \Bf\ $B \propto t^{-5/4}$, while later $B \propto t^{-7/4}$.

Also note that the NS-driven shock never overtakes the freely  expanding   ejecta (radius (\ref{R}) is always smaller than $V_{ej,0} t$). The shock breaks out  hydrodynamically  from the ejecta   into the preexisting wind when ejecta starts to decelerate at (\ref{tslow}), after approximately a month.

\subsection{Propagation of NS wind-driven shock through  pre-explosion wind }
\label{112}

 The Kompaneets approximation (\ref{Komp}),
  in the pre-explosion wind profile (\ref{wind}) takes the form (see Appendix \ref{Sedov} for comparison with Sedov scaling - the resulting relations are similar)
  \be
 \frac{L_w}{4\pi R^2c} =\rho_{w} \left( \partial_t R - v_w \right)^2
 \label{Komp1}
 \ee
  Using the wind power  (\ref{Lw}),   Eq. (\ref{Komp1}) gives
  \be
  R _{PWN} = t v_w+  \sqrt{\frac{L_{w,0} v_w}{ c \dot{M}} } t_\Omega \ln ( 1+ t/t_\Omega)
  \approx  \sqrt{\frac{L_{w,0} v_w}{ c \dot{M}} } t
%  \left\{
%  \begin{array}{cc}
 % \sqrt{\frac{L_{w,0} v_w}{ c \dot{M}} } t & t\ll t_\Omega
%  \nn
  \ee
  where the last relation assumes $t \ll t_\Omega$ and high shock velocity $ V_s\gg v_w$.
  \be
  V_s =  \sqrt{\frac{L_{w,0} v_w}{ c \dot{M}} } = 7\times 10^8 V_{w,8}^{1/2} \dot{m}_{-3}^{-1/2} {\rm cm\, s}^{-1}.
  \label{Vs1}
  \ee

The equipartition \Bf\ is
\be
B\approx  \frac{ \sqrt{2 L_w}}{\sqrt{c} R}= \sqrt{  \frac{2 \dot{M} }{v_w}} \frac{1}{t (1+t/t_\Omega)}
%= 350 \dot{m}_{-3} ^{1/2} V_{w,8}^{-1/2}t_d^{-1}\, {\rm G}.
\label{B2}
\ee
Thus, at later times, for $t \gg t_\Omega$ \Bf\ decreases $\propto t^{-2}$ (one power of time comes from radius  increasing nearly linearly with time, another from decreasing central power).

\subsection{The X-ray continuum and late NIR bump: synchrotron emission from the NS-driven termination shock}

There are two separate components in the early X-ray spectrum:  early, at $t_d \sim 7$, X-ray bump at $\sim 50$ keV and a continuous power-law. The bump disappeared later on, while the continuous component didn't show significant spectral evolution in the 0.3-10 keV  during the 2 months  \cite{2018MNRAS.480L.146R}. In addition, after $\sim $ 40 days there an increase in the NIR emission.

The continuous component is generally consisted with the synchrotron cooled population, resulting in the spectral index $\alpha \approx 0.5$.  We associate the X-ray emission with the particles accelerated at the termination shock emitting in the fast cooling regime.

Suppose the \NS-launched wind is propagating with \Lf\  $\gamma_w$. The peak synchrotron frequency of particles accelerated by the wind termination shock  is then, using the estimates of the \Bf\ (\ref{B1}) and (\ref{B2}),  
\ba &&
\epsilon_s \approx \gamma_w^2 \hbar \frac{e B}{m_e c} =
\left\{
\begin{array} {cc}
\sqrt{2}  \frac{e \hbar}{m_e c^{5/4}}\gamma_w^2  \frac{L_{w,0} M_{ej}^{1/4}}{V_{ej,0}^{3/4}} \left( t^{5/4} \sqrt{1+t/t_\Omega} \right)^{-1} =
50 {\rm keV}\, m_{ej,-2}^{1/4} \gamma_{w,5}^2 \left( t_d^{5/4} \sqrt{1+0.05t_d} \right)^{-1} & \\
\frac{e \hbar}{m_e c} \gamma_w^2 \sqrt{\frac{\dot{M}}{v_w}} \left( {t(1+t/t_\Omega)} \right)^{-1}=
50 {\rm keV}\, \dot{m}_{-3}^{1/2}  v_{w,8} \gamma_{w,5}^2 \left( t_d \sqrt{1+0.05t_d} \right)^{-1} &
\end{array} 
\right.
\label{es}
\ea
where $t_d$ is time since explosion in days and first line corresponds to times when the shock is in the ejecta, while second when it is in the wind.

 The  cooling energy is
 \be
 \epsilon_c \approx \frac{\hbar m_e^5 c^9}{e^7 B^2 t^2}
 \ee
Initially it is very small, well below the injection frequency (\ref{es}). As a result a cooled distribution will form below the injection peak, producing power-law spectrum with $\alpha \approx 0.5$. This is the origin of the persistent component. In the fast cooling regime the particle distribution below the injection peak is independent of the above-the-peak power-law distribution. This explains constant  0.3-10 keV spectrum \citep{2018MNRAS.480L.146R} even for varying luminocity.

Later-on, with \Bf\ in the termination shock decreasing $\propto t^{-2}$, Eq (\ref{B2}), the cooling energy increases sharply with time,
\be
 \epsilon_c \approx \frac{m_e ^5 c^9 \hbar}{e^7} \frac{v_w^{3/2}} {\dot{M}^{3/2} t_\Omega^3} t^4
 \ee
 (It is also a sensitive function of the parameters.). For faster, $v _w \sim 10^9$ cm s$^{-1}$, and lighter, $\dot{M} \sim 10^{-4} M_\odot/$yr, early wind the cooling frequency reaches IR at times $t_d \sim 30 $ days. As a result, injected particles will pile-up at the cooling energy. We suggest this as an origin of the late IR bump.  \footnote {Type-Ia SNe also show IR excess around $\sim 30$ days \citep[\eg][Fig. 3]{2019MNRAS.483..628S}. Given quite different environments, we assume that this similarity is superficial.}

The early X-ray bump should have a somewhat separate origin: it cannot be produced by a constant injection source since in that case the spectrum will be of the broken power-law type ($\alpha = 0.5 $ below the break and $\alpha = p/2 $ above the break), not a spectral bump. We suggest that it is produced by an episode of injection - the estimate  (\ref{es}) for the injection energy early on matches the observed spectral peak.

 \section{Low frequency emission - free-free absorption in the ejecta and the wind}

 At lower frequencies, radio and IR waves can experience free-free absorption  \citep[][Eq. 1.223]{1999acfp.book.....L}  both within the ejecta and in the pre-explosion wind.

Ejecta contribute to free-free absorption a lot
\be
\tau _{ff, ej} = 2 \times 10^{20} m_{ej, -2} ^4 \nu_{GHz}^{-4.2} T_4^{-2.7} t_d^{-10}
\ee
It is optically thin for very high frequencies for a long time
\be
\nu_{GHz} >7 \times 10^4 m_{ej, 0.95} T_4^{-0.64} t_d^{-2.4}
\label{nu}
\ee
Thus, in the radio  and far IR the ejecta remains mostly opaque until the shock breakout from the ejecta, after approximately a month, Eq. (\ref{tslow}).

The free-free optical depth through the wind, with density given by (\ref{wind}),  becomes unity at 
 \be
 r_{ff,wind} = 1.5 \times 10^{16} \dot{m}_{-5} ^{2/3} \nu_{GHz}^{-7/10} T_4^{-9/20} v_{w,8}^{-2/3}
 \label{rff}
 \ee  
%where $\dot{m}_{-5} = \dot{M} /(10^{-5} M_\odot/{\rm yr}$, and $T_4=T/10^4$K.

%THINK MORE HERE.

 The shock (\ref{R}) reaches the optical depth of the order of unity through the pre-explosion wind, Eq (\ref{rff}), for 
 \be
 t = 8 \, \dot{m}_{-5} ^{8/15} \nu_{GHz}^{-14/25} T_4^{-9/25} v_{w,8}^{-8/15}\,m_{ej,-2}^{1/5}{\rm days}
 \ee
 
 The effects of free-free absorption explain  the evolution of the radio and IR luminosities, see \cite[][Fig. 1]{2019ApJ...871...73H}. High frequencies, 341 and 230 GHz, are transparent all along, while lower frequency, 34 $GHz$ traces the expanding radius of the corresponding $\tau=1$ surface.

\section{Population synthesis}
\label{sec:pop_syn}

\subsection{Pre-merger evolutionary channels}

Most calculations of WD-WD mergers are aimed at explaining the Type Ia SNe, thus {\it looking} for detonation \citep[see ][for a recent review]{2014ARA&A..52..107M}.
  Less attention has been given to models that fail to detonate. As we argue, failed SN Ia, that collapse via electron capture, may be related to the FBOTs. \cite{2014MNRAS.438...14D} discussed  the results of the WD-WD mergers and argued  that there is large phase space available for WD-WD mergers to produce an  accretion induced collapse (AIC).  \cite{1985ApJ...297..531N} stressed the role of carbon ignition during WD mergers in order to produce a Type Ia SN. Thus, in order to avoid explosion, there should be little carbon in the system. We suggest then that the primary is a heavy ONeMg WD. In this section we calculate possible evolutionary scenarios and rates for the corresponding mergers.

We use the binary population synthesis (BPS) method to predict the properties of the binary mergers, that is the merger rates, host galaxies and formation channels.  Using  the BPS code \texttt{SeBa} \citep{SPZ96, 2012A&A...546A..70T, Too13}, we simulate the evolution of a large number of binaries following in detail those that lead to the merger of an ONeMg and CO WD. Processes such as wind mass loss, stable \& unstable mass transfer, accretion, angular momentum loss, and gravitational wave emission are taken into account. It was shown by \cite{Too14} that the main source of uncertainty in the BPS outcomes come from the uncertainty in the input assumptions, in particular the CE-phase (CE stands for Common Envelope). For this reason, we follow \cite{2012A&A...546A..70T}, in performing two sets of population synthesis calculations using their model $\maa$ and $\mga$. For full details on the models, see \cite{2012A&A...546A..70T}. 
In short, these models differ from one another with respect to the modeling of the CE-phase. Despite the importance of this phase for the formation of compact binaries and the enormous effort of the community, the CE-phase is still poorly constrained \citep[see][for a review]{2013A&ARv..21...59I}. 

Commonly the CE-phase is modeled in BPS codes by energy conservation \citep{1984ApJ...277..355W}, with a parameter $\alpha$ that describes the efficiency with which orbital energy $ E_{\rm orb}$ is consumed to unbind the CE,  i.e.: 
 $GM_d M_{d, env} / (\lambda R )  = \alpha_{ CE} (E_{orb,init} - E_{orb,final})$,  where $ M_d$ is the mass of the donor star, $M_{d, env} $ the mass of its envelope, $R$ its radius, $\lambda$  the structure parameter of its envelope \citep{1976IAUS...73...75P,1984ApJ...277..355W,1988ApJ...329..764L,1987A&A...183...47D,1990ApJ...358..189D}. This recipe is used in model $\alpha\alpha$ for every CE-phase with $alpha*lambda = 2$ \citep{Nel00}.

 This model is based on a balance of angular momentum with an efficiency parameter $\gamma$ defined as $ (J_{init}-J_{final})/J_{init} = \gamma M_{d, env} / (M_d+M_a)$,  where $J_{init}$ and $J_{final}$ is the angular momentum of the pre- and post-mass transfer binary respectively, and $M_a$ is the mass of the companion.  This model is based on a balance of angular momentum with an efficiency parameter $\gamma = 1.5$ \citep{Nel00}.

\begin{figure}[h!]
\centering
\includegraphics[width=.49\textwidth]{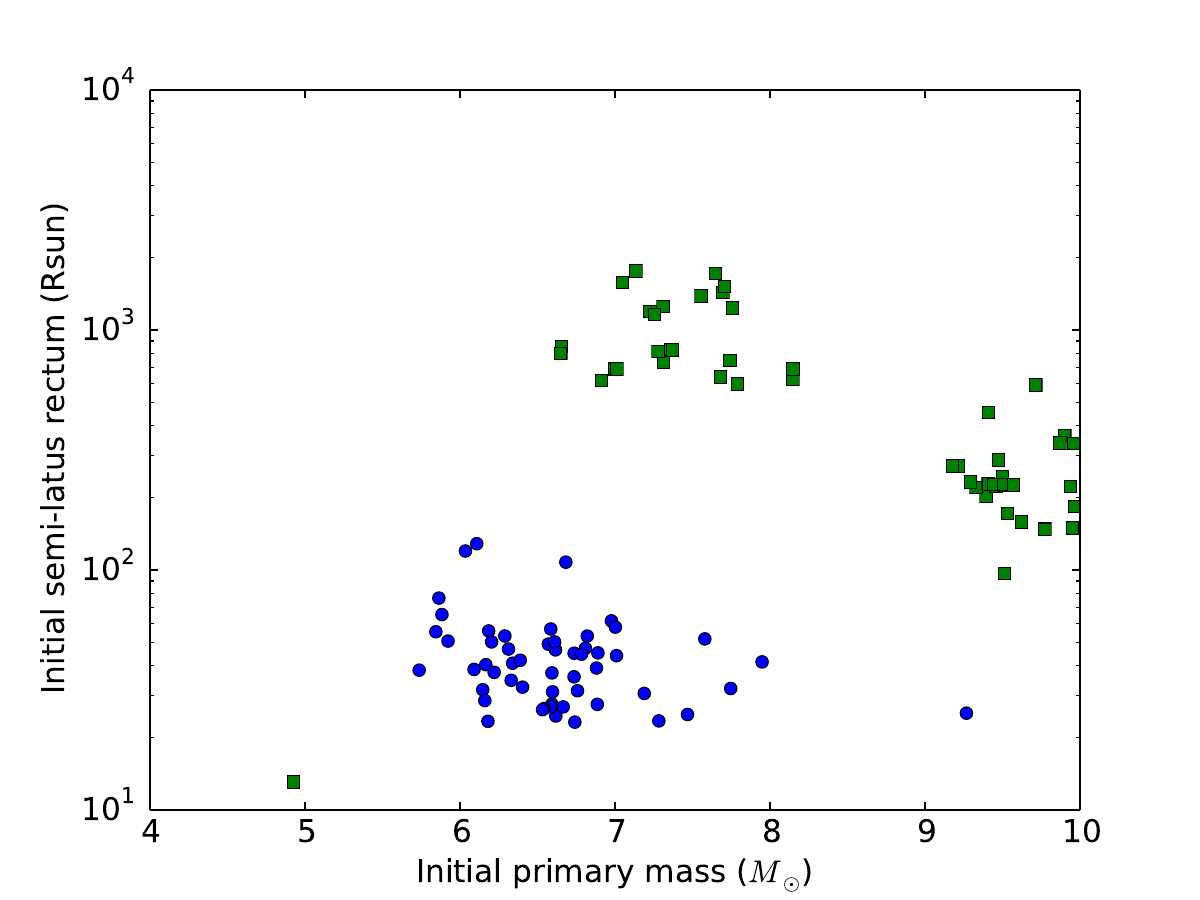}
\includegraphics[width=.49\textwidth]{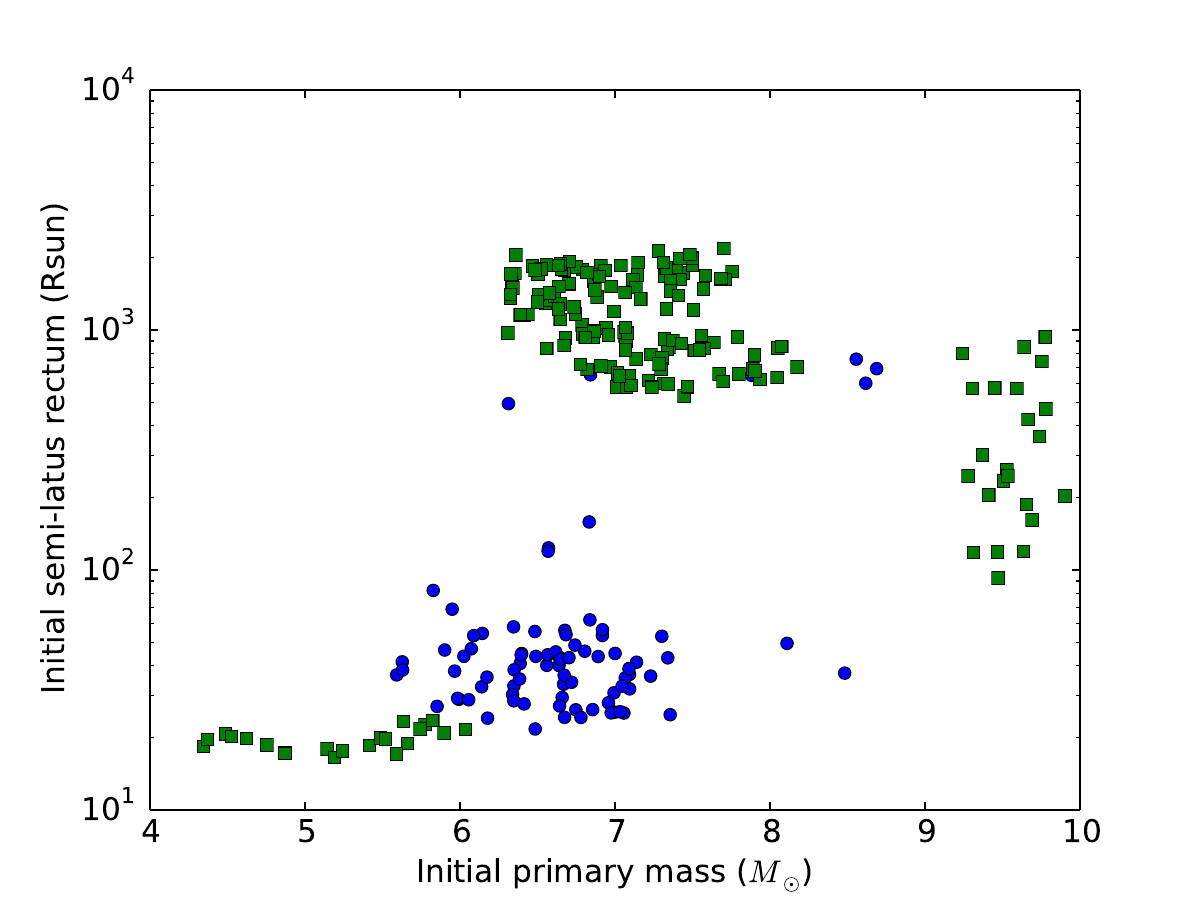}
\caption{Distribution of initial semi-latus rectum for model $\mga$ (left panel) and model $\maa$ (right panel). The color coding is the same as in Fig.\,\ref{minit}. In the case of $\maa$,  green dots at low orbital separations correspond to the systems in which the primary starts with $\sim  6 M_\odot$ and relatively small separation, so that  the first phase of mass transfer is stable. As the secondary accretes, it becomes more massive, its evolution speeds up, and it becomes a ONeMg WD while the primary is still a stripped (hydrogen poor helium-rich) nuclear burning star, which eventually becomes a WD. This is similar to the third evolution channel in \cite{2012A&A...546A..70T}.
 }
\label{ma} 
\end{figure}

\begin{figure}[h!]
\centering
\includegraphics[width=.49\textwidth]{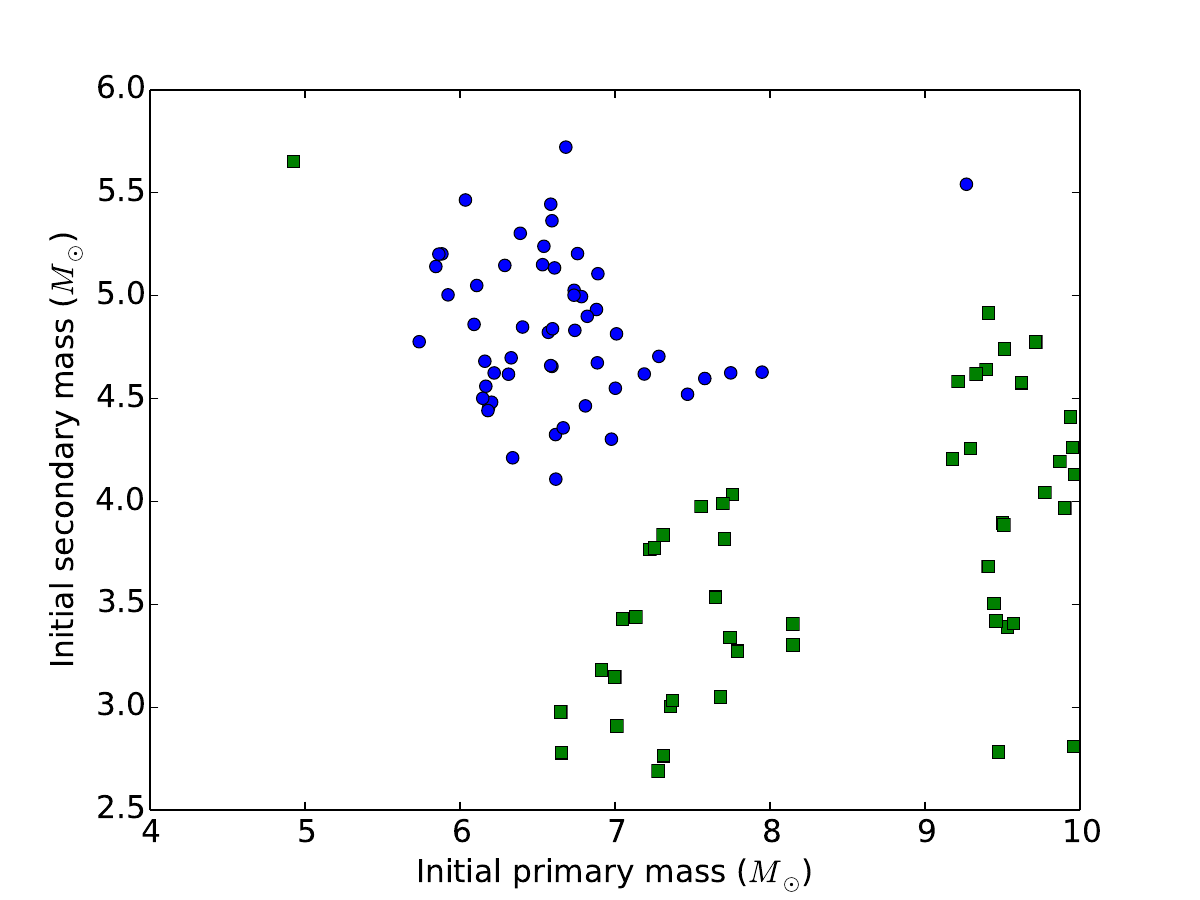}
\includegraphics[width=.49\textwidth]{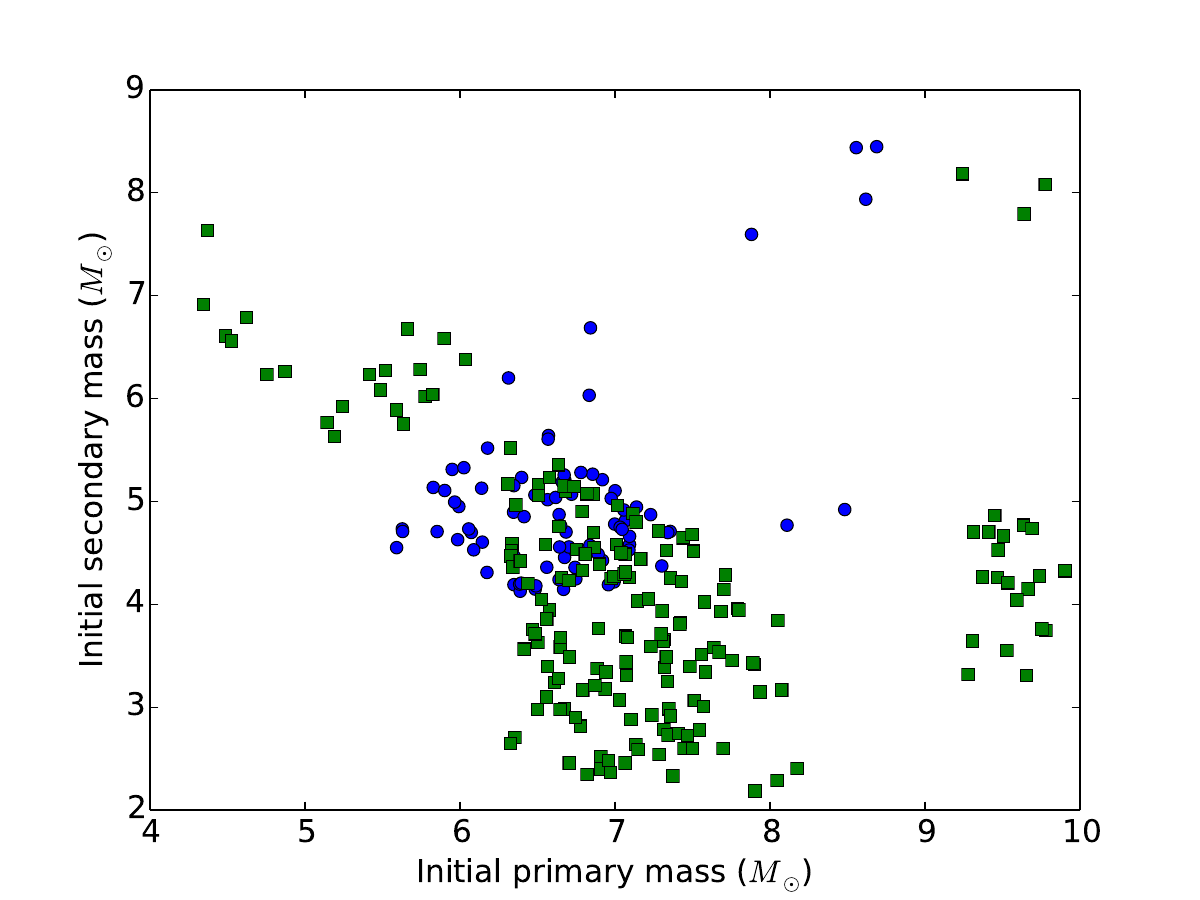}
\caption{Distribution of initial masses for model $\mga$ (left panel) and model $\maa$ (right panel). The primary represents the first formed WD, secondary the last formed WD. With green squares the systems where the ONeMg WD is formed first, with blue circles where this is the last formed WD. In all models the systems marked in blue come from tight orbits where the first phase of mass transfer typically proceeds in a  stable manner. The systems marked in green mostly originate from wider orbits, such that that first phase of mass transfer is likely a common-envelope phase.  }
\label{minit} 
\end{figure}

Figures\,\ref{minit}-\ref{ma} show the initial parameters of binaries leading to mergers between ONeMg and CO WDs in our simulations. Every point represents a single system in the BPS simulations. 
There are different evolutionary paths that can lead to an ONe-CO WD merger, however the two dominant channels consist of systems are: 1) the "inverted channel" (in blue circles) consists of systems with small initial semi-latus rectum $ p\equiv a(1-e^2) < 200 R_\odot$ for which the first phase of mass transfer is stable 2)  systems in the "direct" channel (in green squares) are initially wider and evolve through a common-envelope phase. Typically the ONeMg-WD forms before the companion WD, whereas in the inverted channel the ONeMg WD is formed last (hence the name).  
For single stars, the initial mass  of the progenitor of an ONeMg WD ranges between approximately $6.5-8M_{\odot}$ according to \texttt{SeBa}.This is similar to the range of initial masses in the direct channel where the initially most massive star (i.e. primary) forms the ONeMg WD (majority of green points in Fig.\,\ref{minit}). The progenitors of ONeMg WDs in the "inverted" channel, i.e., the initially less massive star or secondary, denoted in blue, have lower masses initially as these stars accrete a significant amount of mass from their companion stars.

\begin{figure}[h!]
\centering
\includegraphics[width=.49\textwidth]{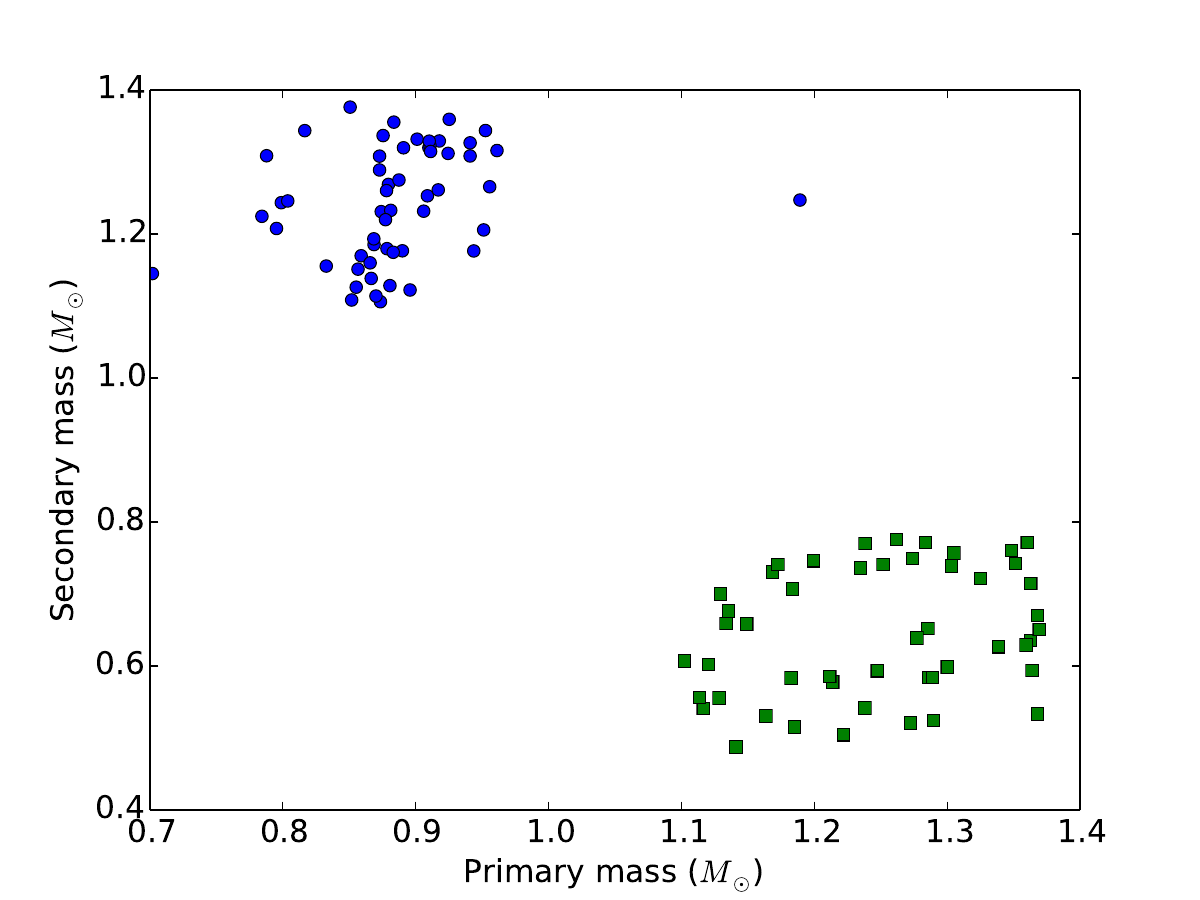}
\includegraphics[width=.49\textwidth]{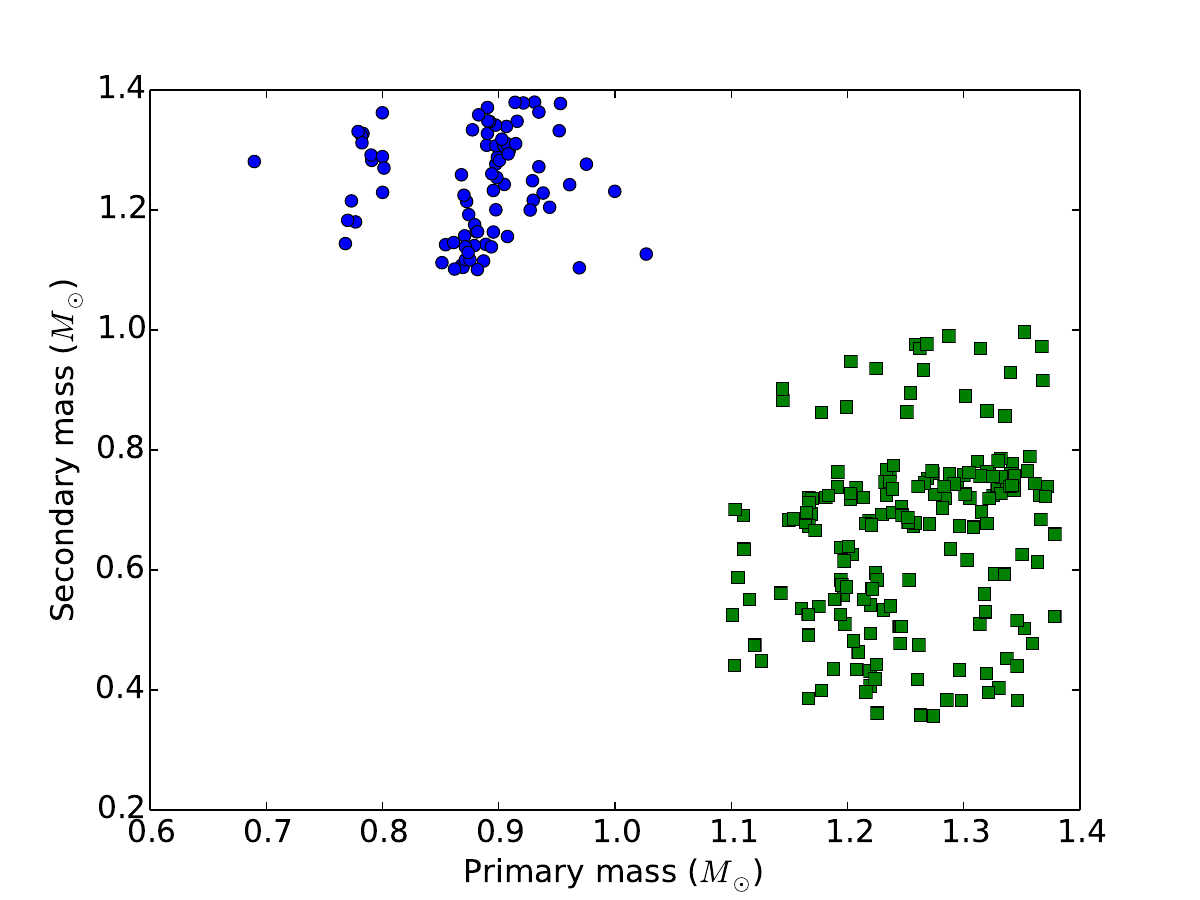}
\caption{Distribution of  WD masses for model $\mga$ (left panel) and model $\maa$ (right panel).  The color coding is the same as in Fig.\,\ref{minit}.  }
\label{mwd} 
\end{figure}

In Fig.\,\ref{mwd}  we show the final masses of the ONeMg and CO WD that merge according to model $\maa$~and~$\mga$ respectively. The masses of the ONeMg WDs are in the range  $1.1-1.4M_{\odot}$, while the majority of CO WDs have masses in the range 0.5-0.8M$_{\odot}$. 
As described above it is possible that the ONeMg WD forms before the other WD in the system (channel 'direct'), or it forms afterwards (channel 'inverted'). In model $\mga$, 48\% of merging ONe-CO DWDs go through the 'direct' channel, whereas for model $\maa$ the fraction goes up to 69\%. The masses of the CO WDs in the 'inverted' channel are systematically higher than those of the 'direct' channel.

\begin{figure}[h!]
\centering
\includegraphics[width=.49\textwidth]{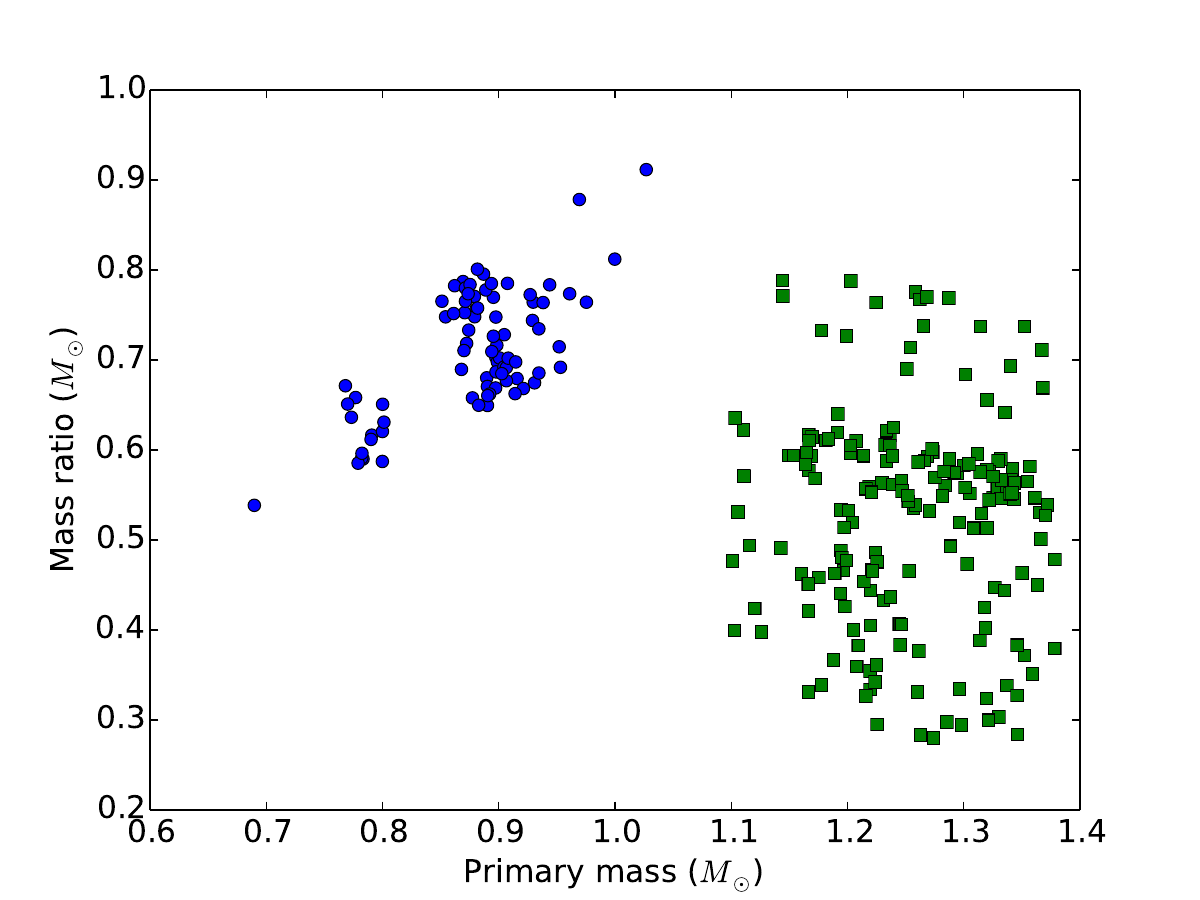}
\includegraphics[width=.49\textwidth]{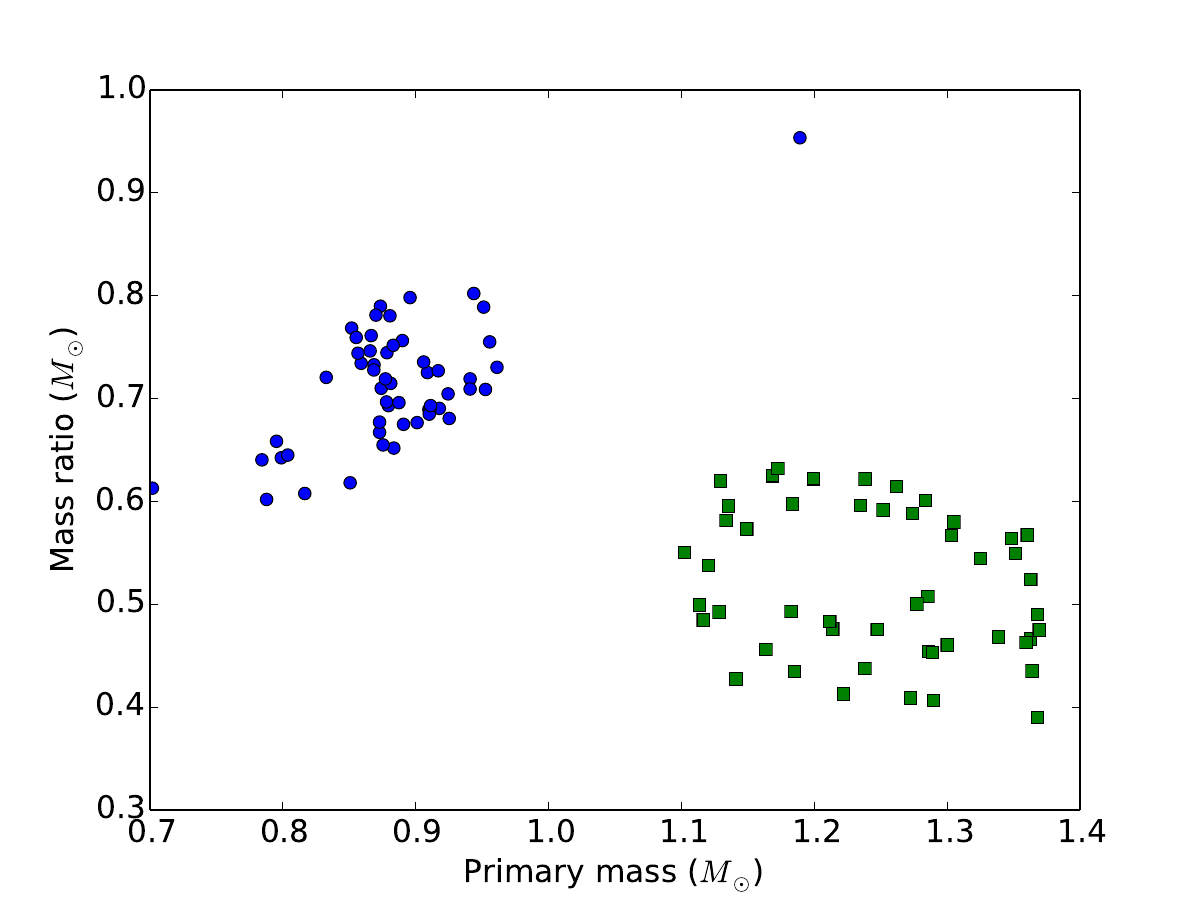}
\caption{Distribution of the mass ratio as a function of the primary mass  for model $\mga$ (left panel) and model $\maa$ (right panel).  The color coding is the same as in Fig.\,\ref{minit}.  }
\label{mwc} 
\end{figure}

In Fig.\,\ref{mwc}  we show the distribution of the mass ratio as a function of the primary mass  at time of merger.  For donor masses in the range $1.1-1.3 M_\odot$  \cite{2004MNRAS.350..113M} (see their Fig. 1) find that mass transfer is always unstable if the companion mass is above  $\sim 0.6$. It is always stable for $\leq 0.2-0.4$. The blue systems are well above the limit for unstable mass transfer. The green systems occupy a larger part of parameter space. The far majority of the systems have a mass ratio that make stable mass transfer unlikely. Also note that given the 'optimistic' stability limits of \cite{2004MNRAS.350..113M} the AM CVn rate is overestimated by orders of magnitude, indicating that mass transfer is likely less stable than the 'optimistic scenario'. 
In addition, the  results from \cite{2004MNRAS.350..113M}  do not take into account the effect of novae outbursts on the evolution of the systems. As shown by  \cite{2012ApJ...748...35S} these outburst have a destabilizing effect on the mass transfer.

%Assuming a constant Galactic  star formation rate of $6 M_{\odot}$ yr$^{-1}$ for the last 10 Gyr\footnote{The assumed star formation history is normalized, such that the total stellar
%mass corresponds to the Galactic stellar mass of $6\times 10^{10} M _{\odot}$yr$^{-1}$ \citep{Too17}}, the current merger rate of CO-ONeMg WDs is $1.9 \times 10^{-4}$yr $^{-1}$ for model $\mga$ and $5.0 \times 10^{-4}$yr $^{-1}$  for model $\maa$. This is still approximately an order of magnitude higher than the estimated short GRB rate. Thus, even among the ONeMg-CO WD mergers, only about $\sim 10\%$ need to produce a short GRB. (This conclusion strongly depends on the assumed beaming of short GRBs.)

%Comparing this to the merger rate of super-Chandrasekhar CO-CO WD mergers, we find that that the synthetic sGRB rate is about 8-16\% of the synthetic SNIa rate and 3-8\% of the observed SNIa rate \citep{2017arXiv170304540M}. 

\subsection{Rates}

Using the BPS simulations, we also estimate the rate of CO-ONeMg WD mergers. Assuming a constant star formation history of $4M_{\odot}$ yr$^{-1}$ for a Hubble time (roughly alike the Milky Way), the current merger rate ranges from  $1.4\times 10^{-4}$yr$^{-1}$ for model $\alpha\gamma$ and $3.4\times 10^{-4}$yr$^{-1}$ for model $\alpha\alpha$. This is in agreement with the BPS calculations of \cite{Yun17} and \cite{2019MNRAS.484..698R}. The CO-ONeMg are less common than mergers between CO-CO WDs for which we find a merger rate of $4.5\times 10^{-3}$yr$^{-1}$. Mergers of CO-CO WDs with a combined mass above Chandrasekhar, the double-degenerate  progenitors model  for supernova type Ia  \citep{1984ApJ...277..355W, Ibe84}, have a merger rate of $1.7-2.2\times 10^{-3}$yr$^{-1}$ in our simulations, about an order of magnitude above that of CO-ONeMg WDs. 

Integrated over time, the total number of CO-ONeMg WD mergers ranging between $(3.5-8.6)\cdot10^{-5}$ per Solar mass of created stars\footnote{This is independent of the assumed star formation history.}. 
The rate of FBOTs has been estimated by  \cite{2014ApJ...794...23D} to be 4\%-7\% of core-collapse supernova. Assuming the core-collapse rate is about 0.0025-0.010M$_{\odot}^{-1}$ \citep{Mao17, Hor11, Gra17}\footnote{The upper limit is calculated by \cite{Mao17}. The lower limit is based on the observed supernova type Ia rate \citep[$(1.3\pm0.1)\dot10^{-3}$M$_{\odot}^{-1}$][]{ Mao17} and the ratio of core-collapse to type Ia supernova \citep[0.25-4][]{Gra17}}, the estimated CO-ONeMg WD merger rate is consistent with the lower limit of the FBOT rate.

%Assuming a constant Galactic  star formation rate of $6 M_{\odot}$ yr$^{-1}$ for the last 10 Gyr \footnote{The assumed star formation history is normalized, such that the total stellar
%mass corresponds to the Galactic stellar mass of $6\times 10^{10} M _{\odot}$yr$^{-1}$ \citep{Too17}}, the current merger rate of CO-ONeMg WDs is $1.9 \times 10^{-4}$yr $^{-1}$ for model $\mga$ and $5.0 \times 10^{-4}$yr $^{-1}$  for model $\maa$ of  \cite{2012A&A...546A..70T}. 
%Consistent with the estimated rates of FBOTs.
 
\subsection{Host galaxies}

\begin{figure}[h!]
\centering
\includegraphics[width=.75\columnwidth]{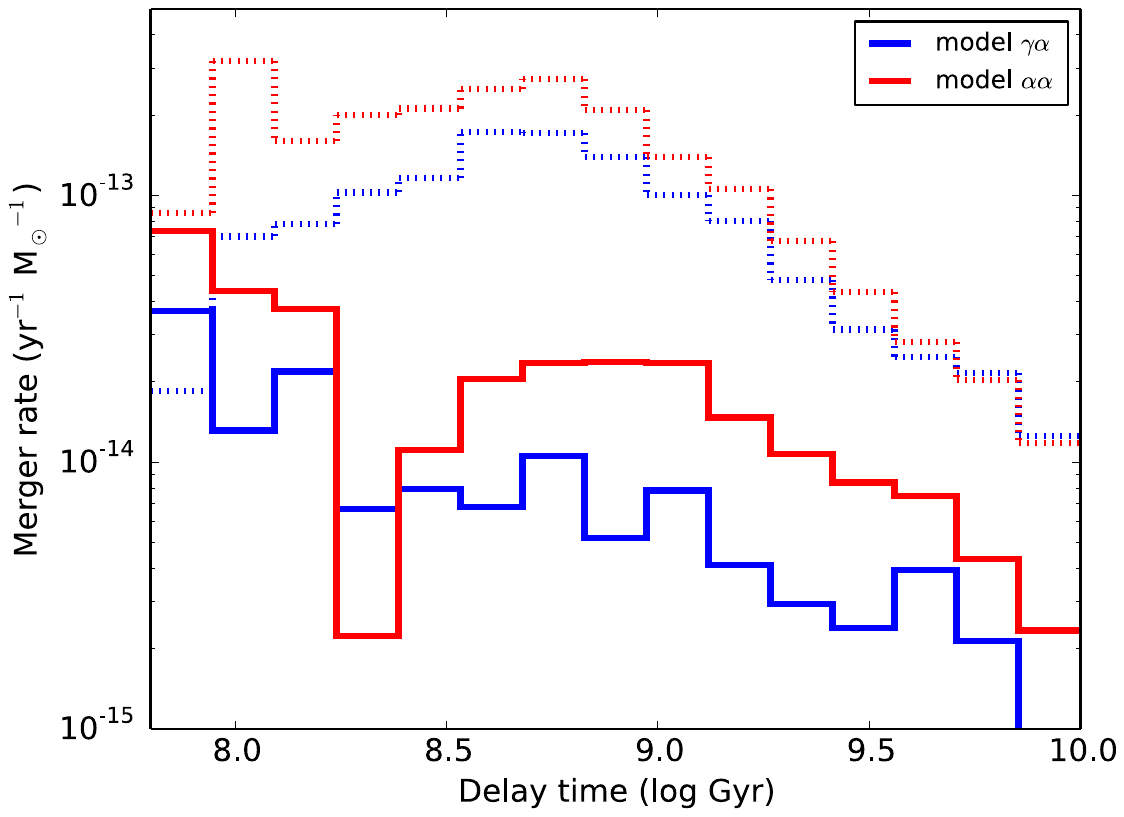}
\caption{Delay time distribution of the CO-ONeMg WD mergers (solid lines) for two models of common-envelope evolution. Model $\mga$ is shown is blue, and model $\maa$ in red. For comparison the super-Chandrasekhar mergers of CO-CO WDs is shown with dotted lines. These typically occur at later delay times than the CO-ONeMg WD mergers.}
\label{DTD} 
\end{figure}

In Fig.\,\ref{DTD} we show the distributions of delay times of the CO-ONeMg mergers after a single burst of starformation. 
Their merger rates peaks at short delay times of about $\sim$ 50-100Myr, with a long tail to long delay times. The peak occurs significantly earlier than expected from the classical type Ia supernovae progenitors consisting of superChandrasekhar mergers of CO-CO WDs \citep[consistent with][]{Yun17, 2019MNRAS.484..698R}. The typical delay time of the CO-ONeMg mergers is closer to that of core-collapse supernovae, which peaks sharply at $\sim$ 50Myr \citep[see e.g. Fig.3 of][]{Zap17}. As a result, we expect the host galaxies of CO-ONeMg mergers to be more similar to those of   core-collapse supernovae instead of Type Ia supernovae. This is consistent with the observed host galaxies of FBOTs \citep{2014ApJ...794...23D}.

%  FBOTs seem to occur in spiral galaxies \citep{2014ApJ...794...23D}. Even though  in this paper we invoke older populations - WDs - we expect that WD  mergers occur preferentially in the galaxies with active star formation - spirals and/or starbursts - since  rates of mergers of compact objects tend to fall of steeply in time after starburst because GW emission is a strong function of orbital separation. Active star formation is needed to produce binaries with small separation. SYLVIA: CITATIONS. Thus, most frequently  WD-WD mergers are expected to occur in spirals/starbursts gal;axies.
%  SYLVIA: CITATIONS.
 % In contrast,  so it hard to explain any systems that preferentially occur in old massive elliptical , like 91bg.  SYLVIA: CITATIONS.

         \section{Alternatives}

An  alternative to the WD merger scenario is the possibility of ultra-stripped envelopes in close binaries \citep{2013ApJ...778L..23T,2015MNRAS.451.2123T}.   Binary interactions
may strip the primary star of the envelope and also affect the mass of the collapsing core \citep{2012ARA&A..50..107L,2004ApJ...612.1044P}.
\cite{2018MNRAS.479.3675M,2018arXiv181105483M}  estimate that   ultra-stripped SNe produce normal slowly rotating pulsars, which are unlikely to produce fast spinning central \NS.  \cite{2018MNRAS.479.3675M} studied the case when the angular momentum is implanted onto the \NS\ only due to  accretion of a tenuous envelope, without accounting for  the progenitor's initial  spins. In contrast, in  the case of  double WD merger the envelope has a lot of angular momentum  and has time to implant it to the core during the shell burning stage.

In a possibly related line of research, \cite{2005PhRvL..94e1102P}, see also \cite{2014MNRAS.438.1005D}, argued that the pulsar J0737-3039B was born through non-standard SN mechanism (presumably via AIC), similar to the   ultra-stripped case considered by  \cite{2015MNRAS.451.2123T}.  \cite{2005PhRvL..94e1102P} also argued for slow initial spin (and slow kick velocity). 
Thus,  ultra-stripped binary cores produce slowly rotating remnants, while AT2018cow needs a powerful central source.

    \section{Discussion}
  
 In this paper we discuss  a channel for  transient emission after electron-capture collapse to a \NS\ following a merger of two WDs. Qualitatively, this channels allows a collapse into a \NS\ to occur with a small envelope mass. As a result the ejecta is light, have high velocity and becomes optically transparent much earlier. This early transparency allows higher radiation efficiency, as energy is not spent on adiabatic expansion of the envelope.  In AT2018cow the ejecta was  the lightest, with only $ \sim 10^{-2} M_\odot$ ejected. In this case the AIC occurred at the time when most of the envelope was already lost to the wind.
 Other FBOTs may have larger remaining envelopes at the moment of AIC and  larger ejected masses, but all smaller than $\sim 0.5 M_\odot$. In our model the envelop mass depends on the mass of the primary ONeMg WD (how close it is to the Chandrasekhar limit), and the mass of the companion (how quickly mass is added to the core, mass loss rate  - how long the shell burning continues). 
  
%  We also advocate the Pulsar Wind paradigm -  the reverse shock, as opposed to the forward shock - as the origin of the X-ray emission, similar to the GRB model of \cite{2017ApJ...835..206L}; see also \cite{Usov92,2006NJPh....8..119L,2011MNRAS.413.2031M}.

  Let us highlight how the   key observational results  discussed by \cite{2019ApJ...872...18M} are explained in our model
  \begin{itemize}
  \item
  A very short rise time to peak, $t_{\rm rise}\sim$ few days - optical transient is generated by an envelope ejected during the bounce from the proto-neutron star. Ejected mass is small, while velocity is nearly relativistic (of the order of the escape velocity from the surface of a proto-neutron star).
  \item Large bolometric peak luminosity - as the ejecta becomes optically thin early on, a large fraction of the internal energy is emitted (as opposed to been spent on adiabatic expansion in conventional SN explosion).
  \item   Persistent blue colors, with lack of evidence for cooling at $\delta t\gtrsim30$ days - later-on the emission starts to become dominated by the non-thermal particles accelerators at the termination shock.
  \item  Large blackbody radius $R_{\rm bb}\sim 8\times 10^{14}$~cm -  wind-driven cavity expands to these scales on time scales of few days, Eq. (\ref{R}).
  \item  Persistent optically thick UV/optical emission with no evidence for transition into a nebular phase - emission is dominated by the radiation-modified forward shock.
  \item Abrupt change of the velocity of the material which dominates the emission  at times $\geq 20$ days - the jet breaks through the ejecta and enters the pre-explosion wind, after time given by (\ref{tslow}).
\item NIR excess of emission - the cooling  energy at the termination shock  moved to IR both due to decreasing \Bf\ in the post-shock flow, Eq, (\ref{es}) and discussion afterwards.
  \item The spectra evolve from a hot, blue, and featureless continuum around the optical peak, to very broad features  - this is a transition from  radiation-modified forward shock  at early  time  to regular matter-dominated forward shock, combined with emission from the termination shock.
  \end{itemize}
In addition
  \begin{itemize}
  \item 
``Late-time optical spectra at $t > 20$ days show line widths of  4000 km s$^{-1}$ ($ 0.01c$, indicating substantially lower outflow velocities than at earlier times (when $v = 0.1$c),
and an abrupt transition from very high velocity to lower velocity emitting material''  \citep{2019ApJ...872...18M}. We associate this transition with the moment when  the NS-driven shock plows through the ejecta and enters the pre-collapse wind, Eq. (\ref{tslow}). This is due to slowing down of the ejecta.
  \item There are indication of  hydrogen  in the spectrum after few weeks \citep{2019ApJ...872...18M}:  if the disrupted WD was of the DA type  one does expect of the order of $10^{-4} M_\odot$ of hydrogen in the pre-collapse wind. This explains the late appearance of hydrogen lines,  presumably when the NS-driven forward shock exits the ejecta and  enters the pre-existing wind. We have no way of telling what the atmosphere of a  post interaction WD will be. However, it is probable that the majority will be a DA, and therefore  the rates would not be wildly different from the ones presented here. Also, in a DA WD hydrogen is limited to a narrow outer layer that will be stripped first during the merger and then mixed up in the envelope.
   \item Similar X-ray and optical luminosities are naturally  explained  as emission from forward and NS wind termination shocks (the latter in the fast cooling regime).
  \item The early X-ray spectral bump is also due to the passage of the peak  frequency (late similar effect  will produce an IR increase)
  \item Erratic inter-day variability of  the X-ray emission \citep{2019ApJ...871...73H} is hard to reproduce within the forward shock scenario \citep{2006NJPh....8..119L}, since the forward shock emission properties depend on the {\it  integrated} quantities (central engine total energy and total matter swept).  But this variability  can be reproduced within the internal shock paradigm:  in the fast cooling regime,  either by the changes of the wind properties of the central source \citep{2017ApJ...835..206L}, minijets in the outflow  \citep{Lyutikov:2006a},  or by Crab flare-like reconnection processes in the shocked pulsar wind \citep{2012MNRAS.426.1374C,2018JPlPh..84b6301L}.
   \item We associate both the high energy as well as radio emission not with the forward shock, as is the case  in regular SNe   \citep{1986ApJ...301..790W,1998ApJ...499..810C}, but with the reverse shock in the newly formed  PWN. As a result,  temporal evolution will be different. 
  \end{itemize}

% There are few issues that do not fall simply with the model. One is that the model would predict hydrogen poor environment, while there are indication of  hydrogen  later-on \citep{2019ApJ...872...18M}, presumably when the NS-driven forward shock enters the pre-existing wind. This is reminiscent of the Type-Ia supernova interacting with the  circumburst  medium   \cite{2013ApJS..207....3S}  - at late times they show H$\alpha$ emission. Presence of hydrogen in the wind is surprising for the double-degenerate scenario for Type-Ia supernovae.

Also note, that AIC with a formation of a neutron star is probably responsible for formation of young pulsars in globular
clusters \citep{1996ApJ...460L..41L}. This is consistent with the present model.

 The present model, connecting   FBOTs to the merger of WDs, is related to the possibility that {\it some} short GRBs come from a similar channel of WD mergers \citep{2017arXiv170902221L}. The detection of gravitational waves associated with a GRB \citep{2017PhRvL.119p1101A} identifies mergers of \NSs\ as  the central engine. It is not clear at the moment whether this identification is generic to the whole class of short GRBs. As discussed by \cite{Lyutikov:2009,2017arXiv170902221L},  there is  a number of observational  contradictions to the binary \NS\ merger paradigm (like extended emission and late flares - both not seen in  GW/GRB170817).   One possibility that is still viable, is that some short GRBs originate from WD mergers. 
  Several  parameters may separate outcomes of WD mergers (\eg\ FBOTs and short GRBs): masses of the merging WDs, the amount of the material lost to the wind,  and the spin right before the AIC (so that AIC can either occur directly to a \NS\ or with a formation of an accretion disk).
 For preferential intrinsic  parameters and viewing angles (\eg\ observer along the spin axis of the collapsing WD), we may see a short GRB, and, later on,  an FBOT.

  %%%%%%%%%%%%%%%%%%%%%%%%%%%%%%%%%%%%%%%%%%%%%%%%%%%%%%%%%%%%%%%%%
\section*{Acknowledgments}
This work had been supported by DoE grant DE-SC0016369 and
NASA grant 80NSSC17K0757.

We  would like to thank Maxim Barkov, Deanne Coppejans, Robert Fisher, Ori Fox, Raffaella Margutti, Danny Milisavljevic, Amir Levinson, Eran Ofek, Amir  Sharon, Liliana Rivera,  Nir Shaviv and   Thomas Tauris for discussions. We also thank Yegor Lyutikov for comments. 

%%%%%%%%%%%%%%%%%%%%%%%%%%%%%%%%%%%%%%%%%%%%%%%%%%%%%%%%%%%%%%%%%

\bibliographystyle{apj}

  \bibliography{/Users/maxim/Home/Research/BibTex}

\appendix 
\section{Sedov approximation for dynamics of pulsar-driven wind propagating through pre-explosion wind }
\label{Sedov}

Another analytical scaling for the dynamics of the strong shock is due to Sedov, which assumes energy conservation (as opposed to momentum flux conservation in case of Kompaneets approximation).  Consider first times shorter than spin-down time $ t \ll t_\Omega$. 
 In the thin shell limit,  the energy in the swept-up shell is the energy deposited into the shell by the central source, $L_0 t$ plus the  kinetic energy of the swept-up matter $M_{swept} v_w^2/2$ where $M_{swept}=4\pi r r_0^2  \rho_0$ is the swept-up mass (assuming cold wind with constant velocity),
 \be
L_0 t+ 4\pi r r_0^2 v_w^2 \rho_0= 4 \pi r r_0^2 \rho_0 v^2/2
\label{E0}
\ee
Dimensionalizing
\ba &&
L_0 =l _ 0 2\pi  r_0^3 v_w^2/2
\nn  && 
t= \tilde{t}( r_0 /v_w)
\nn &&
r = \tilde{r} r_0
\ea
we find
\be
l _ 0  \tilde{t}+  \tilde{r}= \tilde{r} \dot{\tilde{r}}^2
\ee
At early time, neglecting the accumulated energy of the wind, $l _ 0  \tilde{t}\gg   \tilde{r}$
\ba &&
\tilde{r} = l_0^{1/3} \tilde{t}
\nn &&
r=\frac{ {L_0}^{1/3} t}{2 {\pi } ^{1/3}{\rho
   _0} ^{1/3}r_0^{2/3}} = \frac{2^{1/3} L_0^{1/3} v_w^{1/3}}{\dot{M}^{1/3}} t
   \nn &&
   V_{s,S} \approx  \frac{ L_0^{1/3} v_w^{1/3}}{\dot{M}^{1/3}}
   \ea
Comparing with the  Kompaneets approximation, Eq. \ref{Vs1},
   \be
 \frac{ V_s}{V_{s,S} }\approx c^{1/2} \left( \frac{\dot{M} }{L_0 v_w} \right)^{1/6} = 3.6\dot{M}_{-3} ^{1/6} v_{w,8}^{-1/6}
  \label{Vs1}
  \ee
Thus, both Sedov's and Kompaneets's approximations give similar estimates for the shock dynamics at this time.

At times much longer than the spin-down time, $ t \gg t_\Omega$, the central engine has deposited most of it's energy in the shock, so that energy conservation gives
\be
E_0+ 4\pi r r_0^2 v_w^2 \rho_0= 4 \pi r r_0^2 \rho_0 v^2/2
\ee
with $E_0 = L_0 t_\Omega$.
Dimensionalizing 
\be
\epsilon _ 0  = \frac{E_0}{4\pi  r_0^3 v_w^2/2}
\ee
The energy conservation now takes the form
\ba &&
\epsilon _ 0 +  \tilde{r}= \tilde{r} \dot{\tilde{r}}^2
\nn &&
\epsilon _ 0  = \frac{E_0}{4\pi  r_0^3 v_w^2/2}
\ea
with a solution
\be
\tilde{t} = \sqrt{ \tilde{r}(\epsilon _ 0 +  \tilde{r}) }+ \epsilon _ 0 \ln \frac {\sqrt{\epsilon _ 0}}{ \sqrt{ \tilde{r}} + \sqrt{ \epsilon _ 0+ \tilde{r}}}
\ee
At  times when the swept-up kinetic energy is not significant,
\ba &&
\tilde{r} = (3/2)^{2/3} \epsilon _ 0 ^{1/3} \tilde{t} ^{2/3}
\nn &&
r=\frac{3^{2/3} {E_0}^{1/3} t^{2/3}}{2 {\pi } ^{1/3}{\rho
   _0} ^{1/3}r_0^{2/3}} = \frac{3^{2/3}}{2^{1/3}} \frac{E_0^{1/3} v_w^{1/3}}{\dot{M}^{1/3}} t^{2/3}
   \ea
   where the last two relation assume $r \rightarrow 0$ limit.

\end{document}